\documentclass[twocolumn,american,prb,aps,final,superscriptaddress]{revtex4}
\usepackage[T1]{fontenc}
\usepackage[latin9]{inputenc}
\usepackage{color}
\usepackage{amsmath}
\usepackage{graphicx}

\makeatletter
\@ifundefined{textcolor}{}
{%
 \definecolor{BLACK}{gray}{0}
 \definecolor{WHITE}{gray}{1}
 \definecolor{RED}{rgb}{1,0,0}
 \definecolor{GREEN}{rgb}{0,1,0}
 \definecolor{BLUE}{rgb}{0,0,1}
 \definecolor{CYAN}{cmyk}{1,0,0,0}
 \definecolor{MAGENTA}{cmyk}{0,1,0,0}
 \definecolor{YELLOW}{cmyk}{0,0,1,0}
}

\usepackage{stix}
\allowdisplaybreaks

\usepackage{babel}

\makeatother

\usepackage{babel}
\begin{document}
\title{Understanding doped perovskite ferroelectrics with defective dipole
model}
\author{J. Liu}
\affiliation{State Key Laboratory for Mechanical Behavior of Materials, School
of Materials Science and Engineering, Xi'an Jiaotong University,
Xi'an 710049, China}
\author{L. Jin }
\affiliation{Electronic Materials Research Laboratory, Key Laboratory of the Ministry
of Education \& International Center for Dielectric Research, School
of Electronic and Information Engineering, Xi'an Jiaotong University,
Xi'an 710049, China}
\author{Z. Jiang}
\affiliation{School of Microelectronics \& State Key Laboratory for Mechanical
Behavior of Materials, Xi'an Jiaotong University, Xi'an 710049,
China}
\author{L. Liu}
\affiliation{College of Materials Science and Engineering, Guilin Univeristy of
Technology, Guilin 541004, China}
\author{L. Himanen}
\affiliation{Department of Applied Physics, Aalto University, Espoo 00076, Finland }
\author{J. Wei}
\affiliation{Electronic Materials Research Laboratory, Key Laboratory of the Ministry
of Education \& International Center for Dielectric Research, School
of Electronic and Information Engineering, Xi'an Jiaotong University,
Xi'an 710049, China}
\author{N. Zhang }
\affiliation{Electronic Materials Research Laboratory, Key Laboratory of the Ministry
of Education \& International Center for Dielectric Research, School
of Electronic and Information Engineering, Xi'an Jiaotong University,
Xi'an 710049, China}
\author{D. Wang}
\affiliation{School of Microelectronics \& State Key Laboratory for Mechanical
Behavior of Materials, Xi'an Jiaotong University, Xi'an 710049,
China}
\email{dawei.wang@mail.xjtu.edu.cn}

\author{C.-L. Jia}
\affiliation{School of Microelectronics \& State Key Laboratory for Mechanical
Behavior of Materials, Xi'an Jiaotong University, Xi\textquoteright an
710049, China}
\affiliation{\textsuperscript{}Peter Grünberg Institute and Ernst Ruska Center
for Microscopy and Spectroscopy with Electrons, Research Center Jülich,
D-52425 Jülich, Germany}
\date{\today}
\begin{abstract}
While doping is widely used for tuning physical properties of perovskites
in experiments, it remains a challenge to exactly know how doping
achieves the desired effects. Here, we propose an empirical and computationally
tractable model to understand the effects of doping with Fe-doped
BaTiO$_{3}$ as an example. This model assumes that the lattice sites
occupied by Fe ion and its nearest six neighbors lose their ability
to polarize, giving rise to a small cluster of defective dipoles.
Employing this model in Monte-Carlo simulations, many important features
like reduced polarization and the convergence of phase transition
temperatures, which have been observed experimentally in acceptor
doped systems, are successfully obtained. Based on microscopic information
of dipole configurations, we provide insights into the driving forces
behind doping effects and propose that active dipoles, which exist
in proximity to the defective dipoles, can account for experimentally
observed phenomena. Close attention to these dipoles are necessary
to understand and predict doping effects. 
\end{abstract}
\maketitle

\section{Introduction}

For perovskite ferroelectrics, doping chemical elements is an important
way to improve or modify their properties and performances \citep{Jung,Yang,Shi,Limpichaipanit}.
In many cases, it appears that minuscule doping already have strong
effects on the resulting materials. While there is a large amount
of literature on exploiting doping effects experimentally, the nature
and cause of the observed effects are not fully understood, and several
possible factors are proposed to explain experimental results. The
effects of doping induced oxygen vacancies and ferroelectric domains,
among others, are often believed to play important roles \citep{Ang,Renxb,Deng,Zhang_Mn,Hong,Jin,Jin_2,Gao,Chapman,Liu}.
In fact, the exact mechanism of doping effects is hard to identify
by working backward (i.e., to deduce directly from experimental results),
thus a lot of theoretical modeling is needed in this process. Usually,
the most reliable method to calculating doping effect is the first
principle calculation. However, this method has practical difficulty
for minisacle doping where too many atoms are needed in the simulations,
which causes heavy computational burden. Here we focus on the iron
doping BaTiO$_{3}$ and consider this problem in an opposite direction
\citep{generative}. In other words, we first propose a computationally
tractable model regarding how doping works, and then with the effective
Hamiltonian approach \citep{Zhong,Wang_2014,Al-Barakaty,Wang-1},
we explore the consequences of the proposed model. Comparing the simulated
results for samples with and without doping to experimental results,
one will gain better understanding of the nature of doping effects.

Barium titanate (BaTiO$_{3}$), a typical ferroelectric oxide with
perovskite ABO$_{3}$ structure (with Ba on the A site and Ti on the
B site), has been widely investigated due to its high dielectric permittivity,
excellent electrical properties, and environmental friendliness\citep{Senn2016,Qi}.
Doping of BaTiO$_{3}$-based materials with different chemical elements
has been an attractive topic of research \citep{Yu_z,West,Xu,Hennings1982,Jin2016,Yasmm},
with the goal to improve material performance. One important line
of research is to dope rare earth elements in BaTiO$_{3}$. For instance,
Yasmin \textit{et al.} found that the Ce-doped BaTiO$_{3}$ has a
dielectric permittivity as high as 2050 and a decreased Curie temperature,
$T_{C}=313$\,K\citep{Yasmm}. Ganguly \emph{et al.} reported that
at 10\,kHz, the dielectric permittivity for (Ba$_{1-x}$La$_{2x/3}$)TiO$_{3}$
($x=0.1$) can reach 10400 at $T_{C}\simeq168$\,K \citep{Ganguly}.
Ba(Zr$_{x}$Ti$_{1-x}$)O$_{3}$ ceramics, where the B-site Ti is
substituted with Zr, shows enhanced remnant polarization and field
piezoelectric strain coefficient $d_{33}$\citep{Yu_z}, as well as
interesting dielectric properties on the subterahertz frequency range
\citep{Wang-1}. La and Zr co-doped BaTiO$_{3}$ can obtain a dielectric
permittivity as high as 36000 \citep{West}. In addition, doping transitional
metal ions could also achieve novel ferromagnetic properties with
a saturation magnetization value as large as 0.012 emu/g \citep{Xu,Chakraborty}.

In experimental investigations of doped systems, three phenomena are
often observed as the consequences of doping: (i) Strong change of
the hysteresis loop; (ii) Diffused and/or smeared dielectric permittivity
with respect to temperature; (iii) Variations of phase transition
temperatures \citep{Xu,Hennings1982,Weber,Baskaran,Wang,Nanakorn,Kundu,Leontsev,Guo,Huang_0.8}.
Moreover, it is also known from experiments that doping Fe or Mn into
BaTiO$_{3}$ in general makes ferroelectric materials easier to reverse
\citep{Haertling_1999,Chen,Tangsritrakul,Khirade}. Since the doping
effects can be significant even with minuscule doping, the origin
of such effects naturally attracts great scientific attention. Different
explanations have been proposed to understand the mechanism of doping
effects, including (i) Oxygen vacancies and free charges on the doped
lattice point\citep{Jin,Gao,Chapman,Liu,Tangsritrakul,Shuai,scott};
(ii) Local strains \citep{Choi,Wu,Xu_AFM}; (iii) Domains induced
by dopants \citep{Hong,Jin,Jin_2,Wada}. In the present work, we propose
a computationally tractable model where the dipoles associated with
Fe-doped lattice sites and their nearest neighbors are suppressed.
We apply this model to mimic Fe-doped BaTiO$_{3}$ and perform first-principles-based
Monte-Carlo (MC) simulations, showing that the simulated results can
account for many experimentally observed phenomena. Through the aforementioned
approach, we hope to better understand the most important factors
that determine properties of BaTiO$_{3}$ doped with iron.

This paper is organized as follows. In Sec. \ref{sec:Model-and-Method},
we introduce the defective dipole model and the effective Hamiltonian
method for numerical simulation. In Sec. \ref{sec:Doping-effects}
and Sec. \ref{sec:Dipole-distributions}, we apply this model to samples
mimicking Fe-doped BaTiO$_{3}$ and numerically obtain the results
of doping. In Sec. \ref{sec:understanding}, we propose the concept
of active dipoles and use it (along with dipole distributions) to
explain doping effects. Finally, in Sec. \ref{sec:Conclusion}, we
present a brief conclusion.

\section{Method \label{sec:Model-and-Method}}

The transition metal ions we are concerned with, including Fe, Mn
and Co, have one or more vacant orbitals (e.g.,\textsl{ d}-orbitals)
that may host extra electrons. Their electronic properties need a
large on-site energy $U$ to be properly understood \citep{Leontsev,Chen,Anderson,Hubbard}.
When extra charge carriers are (temporarily) captured by these ions,
the localized charges can: (i) distort local electronic band structure
and (ii) introduce extra Coulomb interaction between neighboring sites.
For instance, it was estimated that the interaction energy between
two localized electrons on nearest neighbors can be as large as 2-3\,eV
\citep{Hubbard}. Therefore, it is conceivable that around the dopants,
this type of interaction may strongly affect the dipoles (arising
from displacements of ions) that exist on each unit cell of ferroelectric
materials.

In addition, depending on the impurity energy $\varepsilon_{d}$ (relative
to the Fermi level), the local conductivity could be changed {[}related
to the aforementioned effect (i){]} \citep{Patino}. If the local
conductivity is high, local dipoles cannot exist because the electric
field associated with dipoles will lead to the redistribution of charge
carriers and eventually neutralize the bound charges induced by displacements
of ions, which are responsible for forming the dipoles in the first
place.

Based on the above arguments we propose that dipoles on and around
the Fe sites are suppressed and remain constant. Thus, the number
of suppressed dipole should be equals to 7 times of the doped iron.
The precise number could be calculated by the equation $n_{\textrm{dipole}}=7\times n_{\textrm{Fe}}$,
which is is always satisfied, until two or more defective dipoles
contact with each other. In addition, while we aim at Fe doped BaTiO$_{3}$,
it is likely that similar arguments can be used for other transitional
metal \citep{Zhang_Mn,Chen,Ihrig,Ihrig_valence}.\textcolor{red}{{}
}However, the exact number of defective dipoles induced by one dopant
should be determined empirically. 

We use effective Hamiltonian based MC simulations to obtain finite
temperature properties. A pseudo-cubic supercell of size $\mathrm{12\times12\times12}$
(i.e., 1728 unit cells, 8640 atoms) with periodic boundary conditions
is employed in simulations. Among all the unit cells, we randomly
select a certain number of them to represent Fe ion doped cells. The
dipole moments on these selected sites are set to null in MC simulations.
In addition, due to the influence of the Fe dopants, all its six first
nearest neighbors are set to be defective too. The total energy is
given by the effective Hamiltonian developed in Ref. {[}\onlinecite{Zhong}{]}:

\begin{align}
E^{\textrm{tot}} & =E^{\textrm{self}}\left(\left\{ \boldsymbol{u}\right\} \right)+E^{\textrm{dpl}}\left(\left\{ \boldsymbol{u}\right\} \right)+E^{\textrm{short}}\left(\left\{ \boldsymbol{u}\right\} \right)\nonumber \\
 & +E^{\textrm{elas}}\left(\left\{ \eta_{l}\right\} ,\eta_{H}\right)+E^{\textrm{int}}\left(\left\{ \boldsymbol{u}\right\} ,\left\{ \eta_{l}\right\} ,\eta_{H}\right)
\end{align}
which consists of five parts: (i) the local-mode self-energy, $E^{\textrm{self}}\left(\left\{ \boldsymbol{u}\right\} \right)$;
(ii) the long-range dipole-dipole interaction, $E^{\textrm{dpl}}\left(\left\{ \boldsymbol{u}\right\} \right)$;
(iii) the short-range interaction between soft modes, $E^{\textrm{short}}\left(\left\{ \boldsymbol{u}\right\} \right)$;
(iv) the elastic energy, $E^{\textrm{elas}}\left(\left\{ \eta_{l}\right\} \right)$;
(v) the interaction between the local modes and local strain, $E^{\textrm{int}}\left(\left\{ \boldsymbol{u}\right\} ,\left\{ \eta_{l}\right\} \right)$,
where $\boldsymbol{u}$ is the local soft-mode amplitude vector (directly
proportional to the local polarization) and $\eta_{H}$ ($\eta_{l}$)
is the six-component homogeneous (inhomogeneous) strain tensor in
Voigt notation \citep{Zhong}. The parameters appearing in the effective
Hamiltonian have been reported in Ref. {[}\onlinecite{Nishimatsu}{]}.

In order to understand the doping effects, we build samples of different
dopant concentrations from $0\,\%$ to $2\,\%$, with an increasing
step of 0.2\,\%. For each of the doped samples, we gradually cool
down the system from high (typically 550\,K) down to low (typically
30\,K) temperatures with a step of 10\,K. For each temperature,
we typically carry out 320,000 steps of MC. The first 160,000 steps
are used to equilibrate the system, and the remaining steps to obtain
averaged quantities, e.g., supercell average of local mode. For doped
BaTiO$_{3}$, with increasing dopant concentration, fluctuations close
to phase transitions become so larger that more MC steps (e.g., 640,000)
and/or large supercells (e.g., $18\times18\times18$) are used to
obtain satisfactory results.

To analyze dipole distributions, we have also obtained snapshots of
all the dipoles in the supercell, which are typically captured at
the ending stage of MC simulations when dipoles do not change much.
In addition, averaged dipole configurations are obtained by storing
a snapshot of dipoles every 400 MC steps after the equilibration stage.
We then use all the stored snapshots to calculate the averaged local
dipole in each unit cell and perform statistics on theses dipoles.

Understanding the mechanism of doping has attracted much attention.
Given the variety of doping, fully understanding how doping works
is an immense task and, not surprisingly, different models had been
proposed. In general, depending on the valence state of the dopant
ions, acceptor doping can induce oxygen vacancies, while donor doping
will induce A-site vacancy or conduction electrons \citep{Deng,Ganguly}.
Understanding such complex systems from numerical simulation often
requires \emph{a priori} assumption and modeling about doping. Proposed
mechanisms include: (i) Defective dipoles can be caused by dopants
(which is the mechanism we employ in this work); (ii) Local strain
can be induced due to the different ionic radius of the dopants \citep{Chakraborty};
(iii) As a result of oxygen vacancies and localized charges created
by doping, new dipoles (defect dipoles) can be induced \citep{Liu,XuBx,Shuai,scott};
(iv) Dopants can affect the grain size and/or induce new domains in
the matrix material \citep{Hong,Jin,Jin_2,Wada,Murugaraj}. Recently,
Xu \textit{et al.} show that antiparallel defect dipoles could induce
a negative electrocaloric effect and double-peak behaviors for acceptor
doped BaTiO$_{3}$\citep{XuBx}. Cohen and co-workers successfully
simulated the pinched and shifted hysteresis loop and the large recoverable
electromechanical response using defect dipoles \citep{Chapman,Liu}.
However, due to the long range interaction of dipoles, the fixed dipoles
seem to increase the phase temperature, which is not consistent with
experimental results and verified by our simulations. Therefore, further
refined models or other factors, such as the one discussed here, are
needed.

\section{Doping effects \label{sec:Doping-effects}}

Having described our approach, we now use MC simulations to calculate
basic properties of BaTiO$_{3}$ with different dopant concentration.
We first obtain and show their hysteresis loops, polarizations, as
well as phase transition temperatures, which are summarized to illustrate
the trend of changes with respect to dopant concentration. With all
simulation results, we are also able to obtain the phase diagram of
Fe-doped BaTiO$_{3}$ with respect to temperature and dopant concentration.
In this way, we show that the model can indeed generate important
features of Fe-doped BaTiO$_{3}$ that are observed in experiments.

\subsection{Hysteresis loop}

\begin{figure}[h]
\noindent \centering{}\includegraphics[width=8cm]{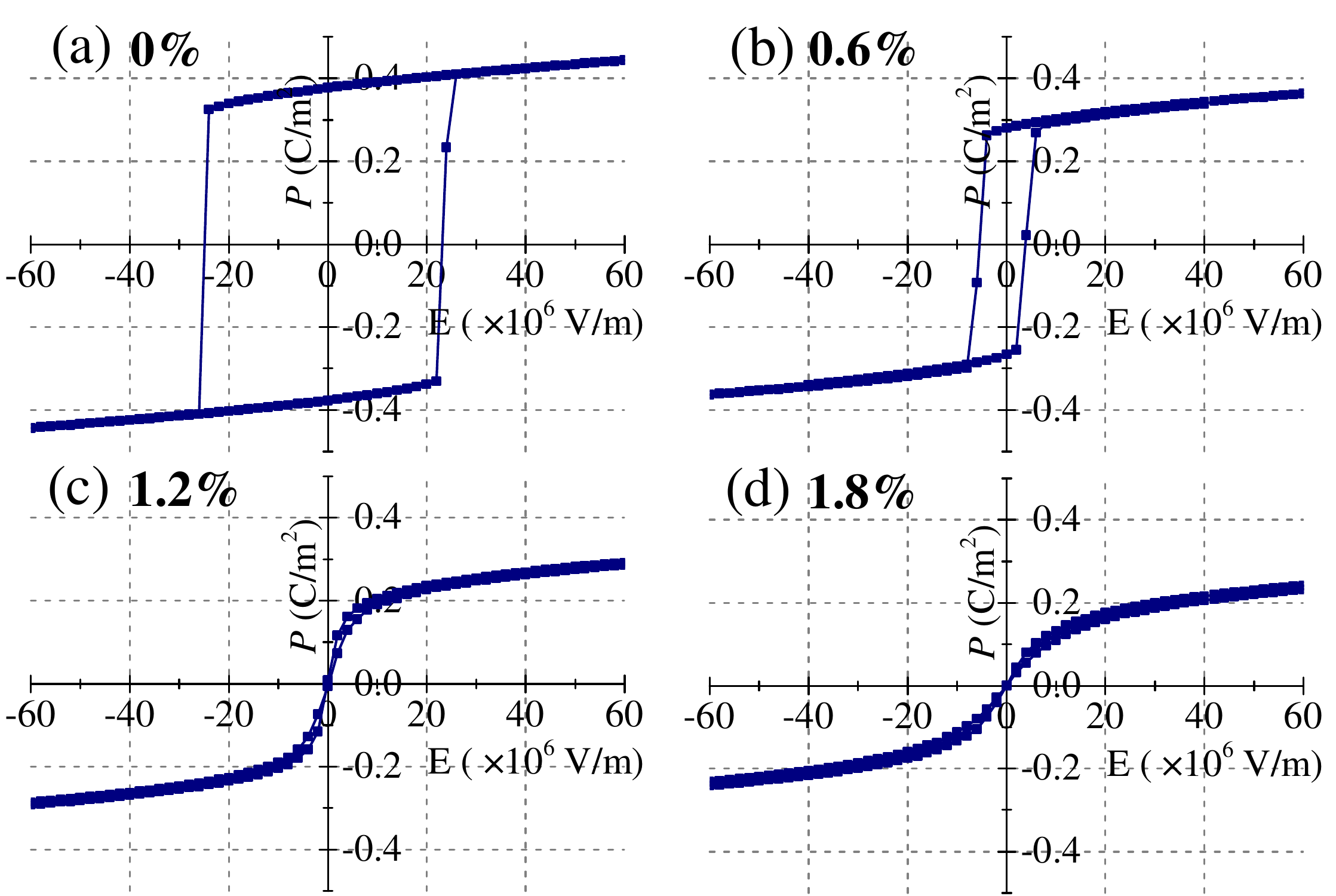}\caption{(a)--(d) Electric hysteresis loops of doped BaTiO$_{3}$ are shown
for various dopant concentration at 300\,K.\label{fig:Hysteresis_loop_doping_percent} }
\end{figure}
Figure \ref{fig:Hysteresis_loop_doping_percent} shows the hysteresis
loops of the doped BaTiO$_{3}$ at selected dopant concentrations
($p=0\,\%,$ $0.6\,\%$, $1.2\,\%$, and $1.8\,\%$) at 300\,K. The
motion of oxygen vacancies is not included, therefore we are not able
to observe the aging phenomenon\citep{Renxb,Zhang_Mn,Gao,Chapman,Liu,Huang_0.8}.
Figure \ref{fig:Hysteresis_loop_doping_percent}(a) demonstrates a
typical hysteresis loop of ferroelectric materials, with large coercive
electric field $E_{c}$ (the electric field when polarization changes
sign) and large remnant polarization $P_{r}$ (the remaining polarization
when the electric field $E=0$\,V/m), as well as large saturation
polarization $P_{s}$ (the polarization at very large $E$). As $p$
increases, both $E_{c}$ and $P_{s}$ decrease. For $p\geq1.2\,\%$,
the hysteresis loop disappears completely {[}see Fig. \ref{fig:Hysteresis_loop_doping_percent}
(c) and (d){]}, indicating some critical changes that will be further
discussed in Sec. \ref{subsec:Hysteresis-loop}. Such changes with
respect to dopant concentration is perhaps the most notable phenomenon
observed and reported in experimental work \citep{Yu_z,Tangsritrakul,Khirade,Qiu,Dutta}.

\begin{figure}[h]
\noindent \centering{}\includegraphics[width=8cm]{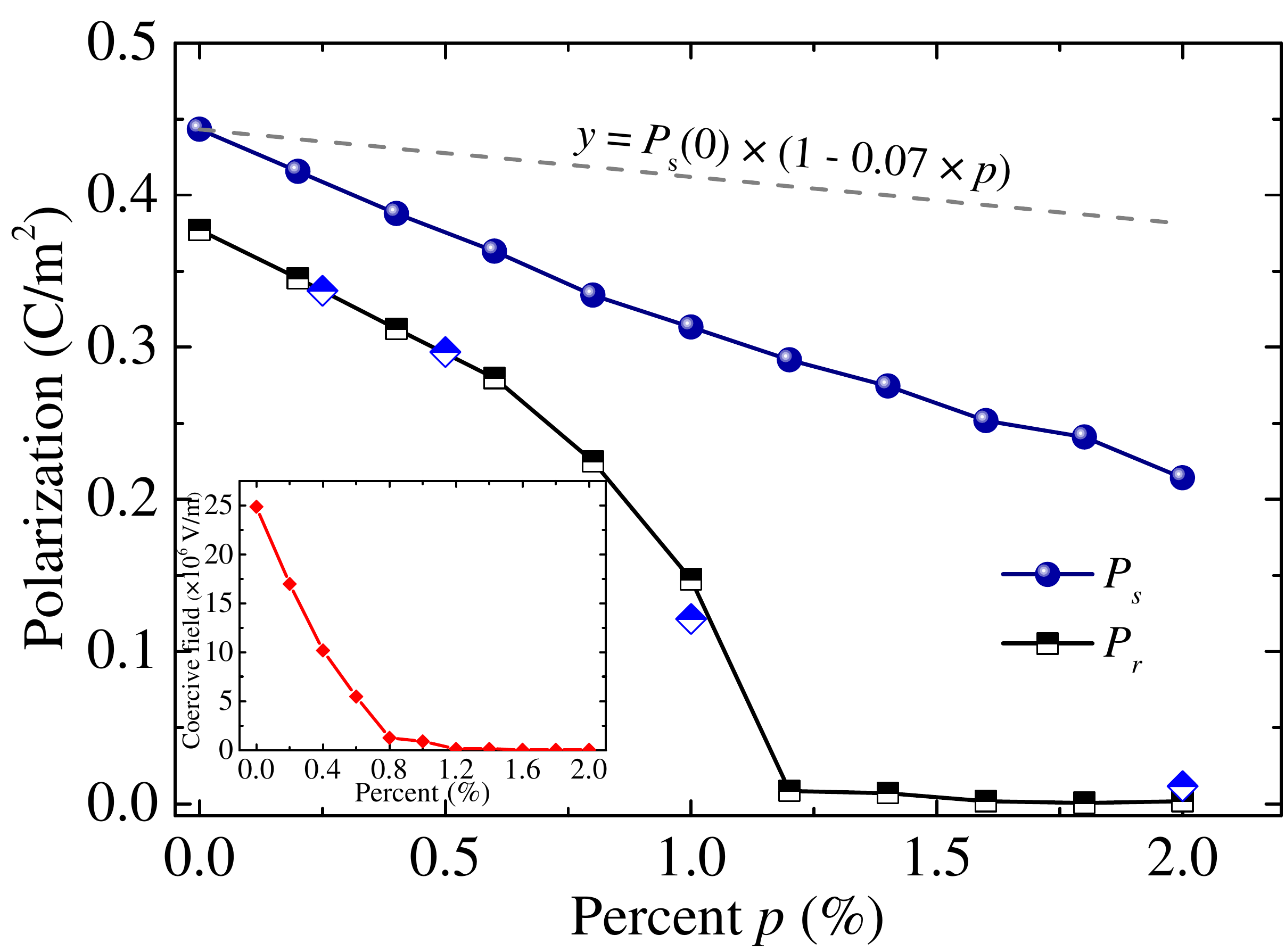}\caption{The saturation polarization, $P_{s}$, the remnant polarization $P_{r}$,
and the coercive field, $E_{c}$ (inset) change with dopant concentration.
(Normalized experimental data from Refs. {[}\onlinecite{Jin2016}{]}
are shown as blue and white diamonds; see the text) \label{fig:Pr_Ec_value} }
\end{figure}

To quantitatively check the doping effects, in Fig. \ref{fig:Pr_Ec_value}
we plot $P_{s}$ (polarization obtained at $E=6\times10^{7}\,\textrm{V/m}$),
$P_{r}$, and $E_{c}$ as functions of $p$, the dopant concentration.
It can be seen that $P_{s}$ depends linearly on the concentration.
According to our assumption, introducing a dopant ion will bring in
seven dead dipoles near the defect sites. Therefore, at the dopant
concentration $p$, the estimated saturation polarization should not
exceed 
\begin{align}
P_{s}\left(p\right)\simeq & P_{s}\left(0\right)\times\left(1-7p\right),\label{eq:sole-reduce}
\end{align}
which is shown as the gray dashed line in Fig. \ref{fig:Pr_Ec_value}.
As a matter of fact, $P_{s}$ declines faster than this estimation,
which indicates the importance of dipole correlations in ferroelectric
materials. Figure \ref{fig:Pr_Ec_value} also shows that both $P_{r}$
and $E_{c}$ have a sudden change at some critical values of $p$
($p\simeq0.8\,\%$ and $p\simeq1.2\,\%$), beyond which they become
zero.

In order to compare to experimental results, we plot the simulation
results of $P_{r}$ to that of Fe-doped 0.5Ba(Zr$_{0.2}$Ti$_{0.8}$)O$_{3}$-0.5(Ba$_{0.7}$Ca$_{0.3}$)TiO$_{3}$
\citep{Jin2016} obtained experimentally in Fig. \ref{fig:Pr_Ec_value}.
The change of $P_{r}$ obtained from our simulations shows some interesting
agreement with experiments \citep{BCZT}. We note that to the experimental
data is normalized to compare with our simulation results. In this
process, we use the polarization of the 0.25\,\% doped sample as
unit (and set this value to that obtained from our simulation) to
scale the values of others. 

\subsection{Phase transition}

\begin{figure}[h]
\begin{centering}
\includegraphics[width=8cm]{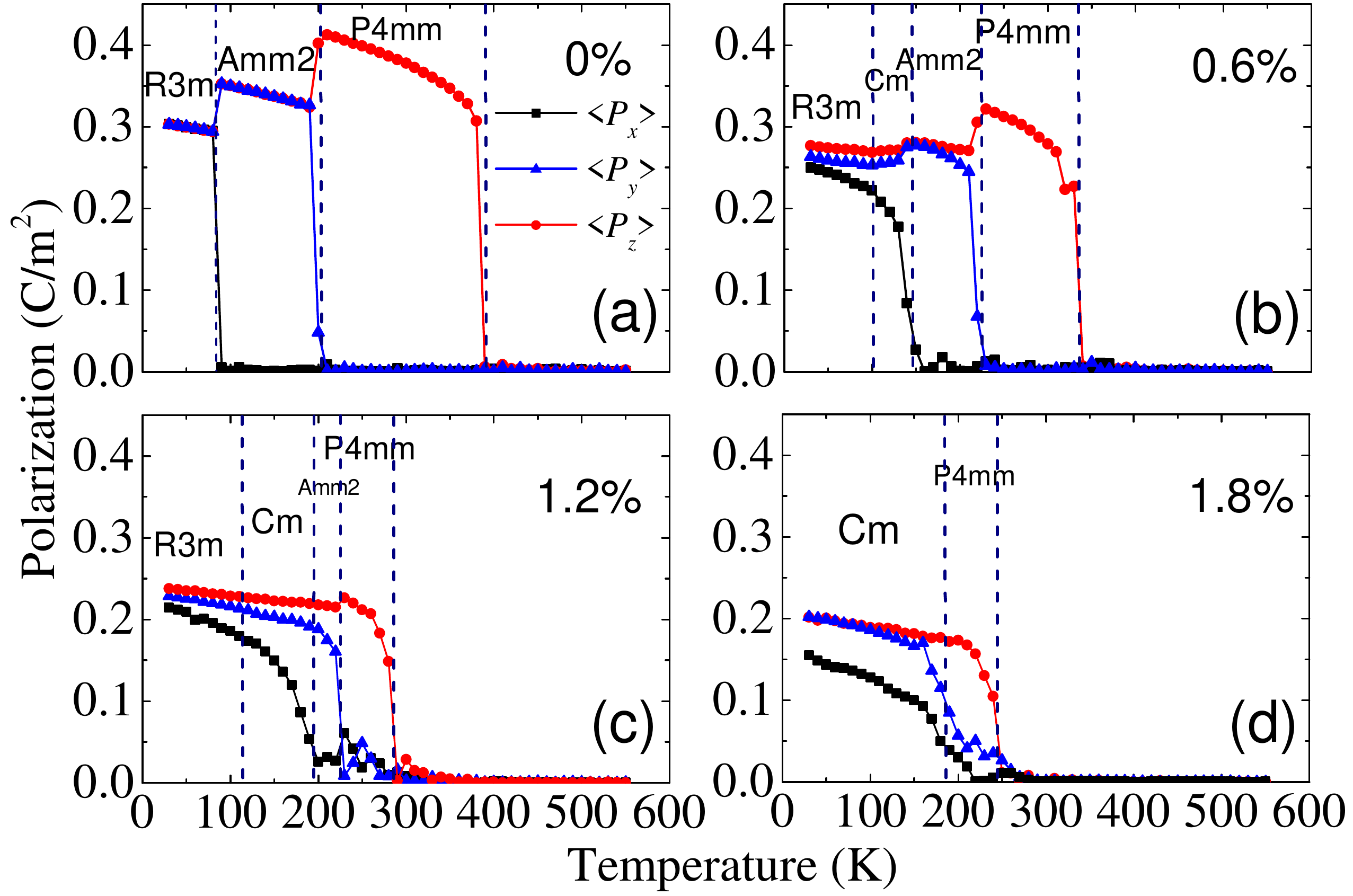} 
\par\end{centering}
\caption{The polarization versus temperature are shown for selected dopant
concentration. The dashed vertical lines show the separation between
different phases. \label{fig:Polarization_vs_Temperature}}
\end{figure}

To reveal how doped BaTiO$_{3}$ evolves with temperature, we show
the averaged components of the polarization$\left\langle \boldsymbol{P}\right\rangle $
in Fig. \ref{fig:Polarization_vs_Temperature}. For pure BaTiO$_{3}$
shown in Fig. \ref{fig:Polarization_vs_Temperature}(a), there exist
four regions: 
\begin{enumerate}
\item For $T>390$\,K, no polarization exists in any direction (i.e., $\left\langle P_{x}\right\rangle =\left\langle P_{y}\right\rangle =\left\langle P_{z}\right\rangle =0$),
showing a paraelectric phase (P); 
\item For $200\,\textrm{K}<T<390$\,K, $\left\langle P_{z}\right\rangle $
becomes nonzero ($0.3-0.42$\,$\textrm{C/\ensuremath{m^{2}}}$),
showing a tetragonal phase (T, space group $P4mm$); 
\item For $90\,\textrm{K}<T<200$\,K, $\left\langle P_{y}\right\rangle $
and and $\left\langle P_{z}\right\rangle $ equal to each other ($\sim0.35$\,$\textrm{C/\ensuremath{m^{2}}}$)
while $\left\langle P_{x}\right\rangle $ is still zero, showing an
orthorhombic phase (O, space group $Amm2$); 
\item For $T<90$\,K, the system reaches the rhombohedral phase (R, space
group $R3m$) with $\left\langle P_{x}\right\rangle =\left\langle P_{y}\right\rangle =\left\langle P_{z}\right\rangle \simeq0.3\thinspace\textrm{C/\ensuremath{m^{2}}}$. 
\end{enumerate}
We note that the above calculated results are consistent with previous
experimental\citep{Huang_0.8,Qiu} and theoretical results \citep{Qi,Vielma,Nishimatsu},
where the polarization being 0.33 (R), 0.36 (O), and 0.27\,$\textrm{C/\ensuremath{m^{2}}}$
(T) \citep{Qi} and the phase transition temperatures are $T_{2}\simeq90$\,K
(O to R), $T_{1}\simeq200$\,K, (T to O) $T_{C}\simeq390$\,K (P
to T)\citep{Nishimatsu}.

For doped BaTiO$_{3}$, Fig. \ref{fig:Polarization_vs_Temperature}(b-d)
show their phase transition temperatures and phase transition sequences.
At $p=0.6\,\%$, all the phases appearing in pure BaTiO$_{3}$ can
still be seen while the phase transition temperatures are different.
On the other hand, when $p=1.2\,\%$, the P-T phase change happens
at $T_{C}\simeq285\,\textrm{K}$ and the T-O at $T_{1}\simeq240\,\textrm{K}$,
leaving a narrower region for the T phase. Finally, at $T_{2}=180\,\textrm{K}$,
the system changes from the $Amm2$ phase (orthorhombic) to the $Cm$
phase (monoclinic, M). At $p=1.8\,\%$, the phase transitions become
completely different with the $Cm$ phase appearing at the lowest
temperatures. In general, it can be seen from Fig. \ref{fig:Polarization_vs_Temperature}
that $\left\langle \boldsymbol{P}\right\rangle $ decreases with $p$.
In addition, for large $p$, we have to endure some ambiguity in determining
the phase transition temperatures as the phase transition become diffused.

\begin{figure}[h]
\begin{centering}
\includegraphics[width=8cm]{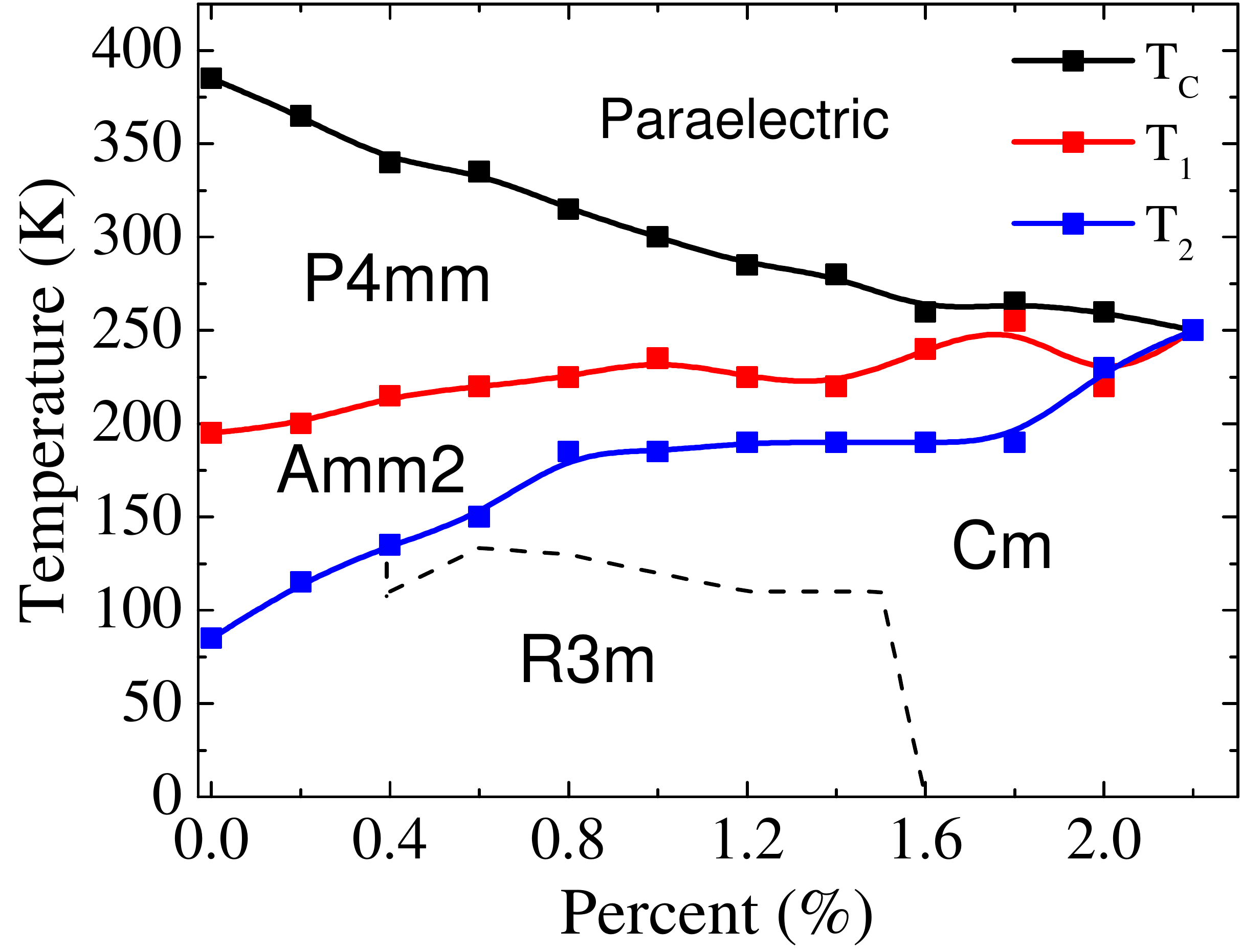} 
\par\end{centering}
\caption{Phase transition temperatures are plotted against dopant concentration.\label{fig:phase_diagram} }
\end{figure}
Doping effects on phase transitions can be summarized in the phase
diagram shown in Fig. \ref{fig:phase_diagram}, which reveals that
for Fe-doped BaTiO$_{3}$ with $p\leq2.0\,\%$, four phases exist:
(i) Between $T_{C}$ and $T_{1}$, the $P4mm$ phase; (ii) Between
$T_{1}$ and $T_{2}$, the $Amm2$ phase; (iii) Below $T_{2}$, the
$R3m$ or $Cm$ phase. The existence of the $R3m$ and the $Cm$ phase
indicates a possible morphotropic boundary that is interesting for
potential performance enhancement \citep{boundary}. We can also see,
as $p$ increases, $T_{C}$ decreases quickly from 375\,K (at $p=0.0\,\%$)
to \textasciitilde 250\,K (at $p=2.0\,\%$), resulting in a decreasing
rate of 57\,K/(1\% doping). Such decreasing rate is much smaller
for Mn (with experimental results being \textasciitilde 15\,K/(1\%
doping) \citep{Chen}, and other similar elements \citep{Ihrig,Ihrig_valence}.
On the other hand, $T_{2}$ increases from 85\,K (at $p=0\,\%$)
to \textasciitilde 250\,K (at $p=2.0\,\%$). Moreover, the shift
of $T_{1}$ is relatively small comparing to the other two. The convergence
of $T_{C}$, $T_{1}$ and $T_{2}$ around $p\simeq2.0\,\%$ agree
very well with experimental results with doped BaTiO$_{3}$ \citep{Ihrig}
, which also slightly resembles what happens with Ba(Zr,Ti)O$_{3}$
when Zr concentration increases\citep{Yu_z,Yu_BZT,Kuang_BZT}. Both
the variation trend and speed of character temperature will be further
discussed in Sec. \ref{subsec:Phase-transition-temperature}.

\section{Dipole distribution \label{sec:Dipole-distributions}}

\noindent In order to understand doping effects on macroscopic properties
of BaTiO$_{3}$, we need to know how microscopic dipole configuration
responds to doping. To this end, in this section we focus on orientation
and magnitude distributions of dipoles in pure and doped BaTiO$_{3}$
to gain insights into how doping works.

\subsection{Orientation \label{subsec:Dipole-orientation-distribution}}

\begin{figure}[h]
\begin{centering}
\includegraphics[width=8cm]{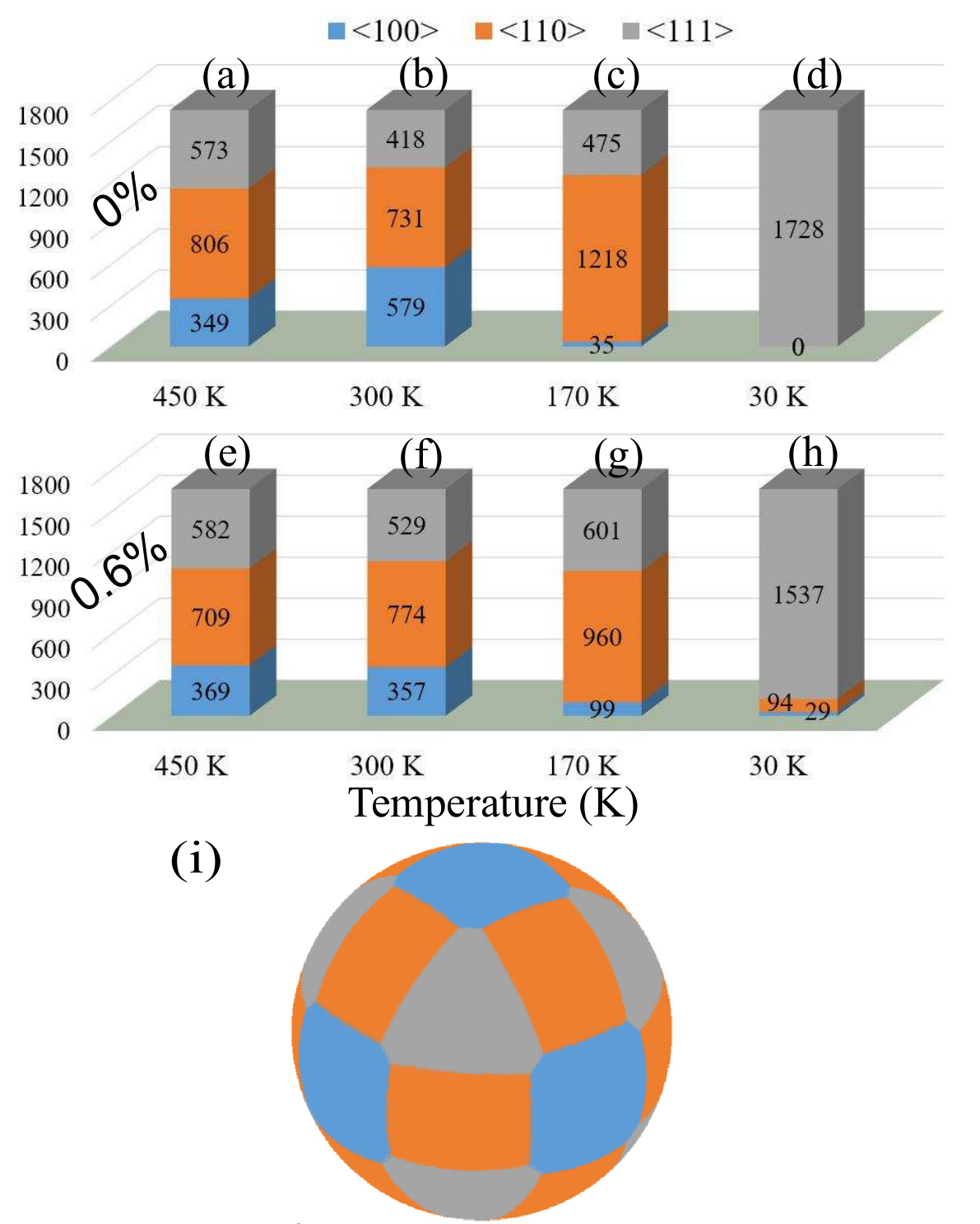} 
\par\end{centering}
\caption{Dipole orientation distribution calculated from dipole snapshot at
450\,K, 300\,K, 170\,K, and 30\,K. (a-d) Pure BaTiO$_{3}$; (e-h)
Doped BaTiO$_{3}$ with $p=0.6\,\%$; (i) Dipole orientation distribution
on a unit sphere. \label{fig:Direction_analysis} }
\end{figure}
In Fig. \ref{fig:Direction_analysis}, the orientation distribution
of dipoles for pure and $0.6\,\%$ doped BaTiO$_{3}$ are shown for
selected temperatures (450, 300, 170, and 30\,K). At each temperature,
we categorize dipoles as $\left\langle 100\right\rangle $, $\left\langle 110\right\rangle $,
or $\left\langle 111\right\rangle $ type dipoles, depending on their
orientations. Figure \ref{fig:Direction_analysis}(i) shows a schematic
drawing of how the categorization is performed and Fig. \ref{fig:Direction_analysis}
(a-h) show the number of dipoles in each category for pure BaTiO$_{3}$
at the selected temperatures.

Figure \ref{fig:Direction_analysis}(a) shows that, at\textbf{ }$T=450$\,K
(paraelectric phase), owing to the large thermal fluctuation, all
three types have significant number of dipoles, with 20.2\,\% for
$\left\langle 100\right\rangle $, 46.6\,\% for $\left\langle 110\right\rangle $,
and $33.2\,\%$ for $\left\langle 111\right\rangle $. The 0.6\,\%
doped BaTiO$_{3}$ {[}Fig. \ref{fig:Direction_analysis}(e){]} almost
has the same distribution. Given such a distribution, at this temperature
the averaged values (e.g. $\left\langle \boldsymbol{P}\right\rangle $)
are very small (see Fig. \ref{fig:Polarization_vs_Temperature}),
producing a paraelectric phase.

At $T=300$\,K (macroscopic $P4mm$ phase), Fig. \ref{fig:Direction_analysis}(b)
shows that the number of $\left\langle 100\right\rangle $ dipoles
have significantly increased, while the number in the other two categories
decreased. Besides, snapshots of dipole configurations reveal that
most dipoles ($>95\,\%$) has positive $z$ component for both pure
and 0.6\% doped BaTiO$_{3}$. This means that a preferred direction
($\left[001\right]$ here) has established at $300$\,K in the system,
likely due to the longe-range dipole-dipole interaction. At $T=170$\,K
{[}$Amm2$, see Fig. \ref{fig:Direction_analysis}(c) and (g){]},
the averaged polarization $\left\langle \boldsymbol{P}\right\rangle $
is along the $\left[011\right]$ direction, consistent with experimental
results. However, for a single snapshot, a large portion (\textasciitilde 27\,\%)
of dipoles belongs to the $\left\langle 111\right\rangle $ category.
In addition, doping gives rise to more ``disobedient'' dipoles that
do not follow the overall orientation. For instance, in pure BaTiO$_{3}$
70.5\,\% of the dipoles belong to the $\left\langle 110\right\rangle $
category, while in the 0.6\,\% doped BaTiO$_{3}$ the number of dipoles
is smaller (57.8\,\%). At\textbf{ }$T=30$\,K ($R3m$ phase), Fig.
\ref{fig:Direction_analysis} (d) shows that all the dipoles in BaTiO$_{3}$
belong to the $\left\langle 111\right\rangle $ category. In fact,
all the dipoles point to a particular one of the $\left\langle 111\right\rangle $
directions, making the system the $R3m$ phase. In contrast, the 0.6\,\%
doped BaTiO$_{3}$ again have some ``disobedient'' dipoles belonging
to the other two types {[}Fig. \ref{fig:Direction_analysis} (h){]}.
In general, Fig. \ref{fig:Direction_analysis} shows that the number
of dominant dipoles decreases with doping.

Below $T_{C}$, the system includes three phases, $R3m$ ($Cm$),
$Amm2$, and $P4mm$, with the average polarization ($\left\langle \boldsymbol{P}\right\rangle $)
pointing along $\left\langle 111\right\rangle $, $\left\langle 110\right\rangle $,
and $\left\langle 100\right\rangle $, respectively. The \emph{local
dipole structure} information associated with Fig. \ref{fig:Direction_analysis}
provides some insights about the phase transitions shown in Figs.
\ref{fig:Polarization_vs_Temperature} and \ref{fig:phase_diagram}.
It can be deduced that: (i) The P to T phase transition is mostly
an order-disorder phase transition since the local dipole orientation
distribution remain similar (i.e., no major dipole rotation happens)
while a ferroelectric phase establishes when the temperature drops
from 450\,K to 300\,K. Therefore the macroscopic phase transition
happens mostly due to the correlation length increase of local dipoles;
(ii) The T to O and O to R phase transition are mixtures of displacive
and order-disorder type. The complementary changes of the $\left\langle 100\right\rangle $
and $\left\langle 110\right\rangle $ dipoles mark the orientation
conversion from the $\left\langle 100\right\rangle $ type to the
$\left\langle 110\right\rangle $ type dipoles (for O to R, it is
the $\left\langle 110\right\rangle $ to $\left\langle 111\right\rangle $
conversion). At the same time, the presence of the $\left\langle 111\right\rangle $
dipoles in all four temperatures indicates the existence of uncorrelated
local rhombohedral regions, which eventually become correlated (via
order-disorder phase transition) at low temperature to form the long-range
rhombohedral phase {[}see Fig. \ref{fig:Direction_analysis}(c,d)
and (g,h){]}. Such observations are critical to understand how $T_{C}$,
$T_{1}$, and $T_{2}$ change with respect to doping.

\subsection{Magnitude}

Since the distribution of dipole components ($u_{x}$,$u_{y}$ and
$u_{z}$) can shed more light on how doping works, we will also analyze
the distributions of dipole magnitudes at different temperatures and
pay particular attention to their evolution with dopant concentration.

\begin{figure*}[!t]
\noindent \begin{centering}
\includegraphics[width=12cm]{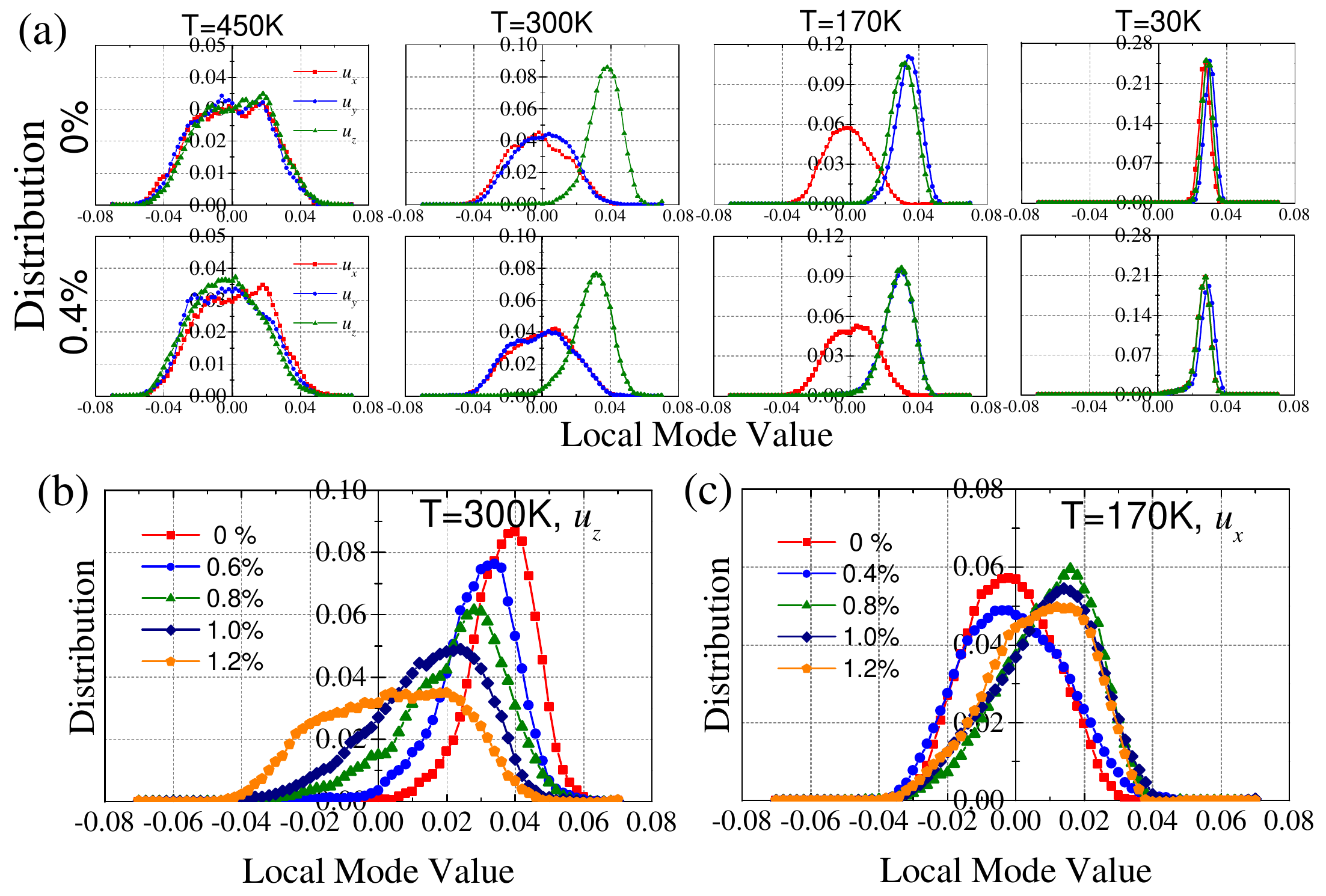} 
\par\end{centering}
\caption{(a) The distributions of dipoles for pure BaTiO$_{3}$ (first row)
and 0.4\% doped BaTiO$_{3}$ (second row) at selected temperatures;
(b) \textit{$u_{z}$} distribution for different doping concentration
at 300\,K; (c) \textit{$u_{x}$} distribution for different doping
concentration at 170\,K. All dipole distributions are normalized.
\label{fig:Ti_displacement} }
\end{figure*}
Figure \ref{fig:Ti_displacement}(a) compares the distributions of
pure BaTiO$_{3}$ to that of the 0.4\,\% doped one. At 450\,K, both
samples have nearly symmetric distribution for $u_{x,y,z}$ with respect
to $\mathbf{u}=0$, which results in the paraelectric phase ($\left\langle \mathbf{u}\right\rangle =0$)
when averaging is performed. At 300\,K, for both samples, $u_{z}$
deviates from the symmetric distribution around zero and centers around
$0.038$. Owing to this deviation, the averaged local mode $\left\langle u_{z}\right\rangle $
is no longer zero and the sample displays a $[001]$ polarization
(T phase). Careful inspection shows that $u_{z}$ of the $0.4\,\%$
doped BaTiO$_{3}$ has a slightly broader peak, which is shifted toward
0 (centering around $0.032$). Similar phenomena are also observed
at 170\,K and 30\,K. In general, comparing to pure BaTiO$_{3}$,
the $0.4\,\%$ sample has its characteristic peak slightly modified,
becoming broader and lower, and its center moving towards zero. It
can be concluded that a small amount of ``disobedient'' dipoles
must exist that have slightly changed these distributions. These ``disobedient''
dipoles can also be used to account for the increased number of minority
dipoles seen in Fig. \ref{fig:Direction_analysis} (e-h). Such dipoles
will be further discussed in Sec. \ref{subsec:Active-dipoles}. Moreover,
we note that the results for pure BaTiO$_{3}$ presented here are
consistent with Ref. {[}\onlinecite{Qi}{]}.

More details are shown in Figs. \ref{fig:Ti_displacement} (b) and
(c). At 300\,K {[}Fig. \ref{fig:Ti_displacement} (b){]}, the distribution
of $u_{z}$ changes strongly with doping, following the aforementioned
trend. At $p=1.0\,\%$, the\textbf{ }peak is already lower by \textasciitilde 50\,\%
than the initial peak, and the center of the peak has changed from
$\sim0.038$ to $\sim0.02$. When $p=1.2\,\%$, the distribution almost
becomes symmetric around zero, giving rise to a paraelectric phase,
which is consistent with Fig. \ref{fig:Polarization_vs_Temperature}(c).
At 170\,K {[}Fig. \ref{fig:Ti_displacement} (c){]}, while $u_{x,y}$
have peaks of nonzero values, $u_{z}$ starts from a symmetric distribution
around zero and develops a peak centering around 0.018 at $p=0.8\,\%$
(which changes the system from the $Amm2$ to the $Cm$ phase), but
the peak shifts toward zero upon further increase of doping concentration.
This feature is reflected by the $Amm2$ and $Cm$ phase boundary
shown in Fig. \ref{fig:phase_diagram}.

\section{Understanding doping effects \label{sec:understanding}}

In order to identify the mechanism of doping, we also examine how
dipole configurations evolve during MC simulations. Our results reveal
that, in doped BaTiO$_{3}$, some dipoles become more active and tend
to change a lot during simulations even far away from phase transition
temperature. In this section, we will first introduce the concept
of ``active dipole'', and then use it along with dipole distributions
to understand doping effects shown in Sec. \ref{sec:Doping-effects}.

\subsection{Active dipoles \label{subsec:Active-dipoles}}

Active dipoles are those dipoles that can substantially change its
states during MC simulations. Unlike the majority of dipoles, which
often fluctuate around their equilibrium position and determine the
macroscopic polarization, active dipoles can easily rotate their direction
and change their magnitudes, while such rotations are forbidden for
most dipoles. In other words, the most important quality of active
dipoles is that they are much more active than other dipoles even
at temperatures much lower than the phase transition temperature.

Practically, we identify active dipoles with the following procedure:
(i) We find the average polarization ($\left\langle \boldsymbol{P}\right\rangle $)
at a given temperature; (ii) Choose a snapshot from the MC simulation
and find dipoles pointing along directions that are different from
$\left\langle \boldsymbol{P}\right\rangle $. According to our simulation,
these dipoles are special in that: (i) They have relatively fixed
locations; (ii) Their quantity is very small (less than $3.3\,\%$
for Fe-doping BaTiO$_{3}$ at $p=2.0\,\%$); (iii) Their directions
change from time to time, making experimental detection hard.

\begin{figure}[h]
\begin{centering}
\includegraphics[width=8cm]{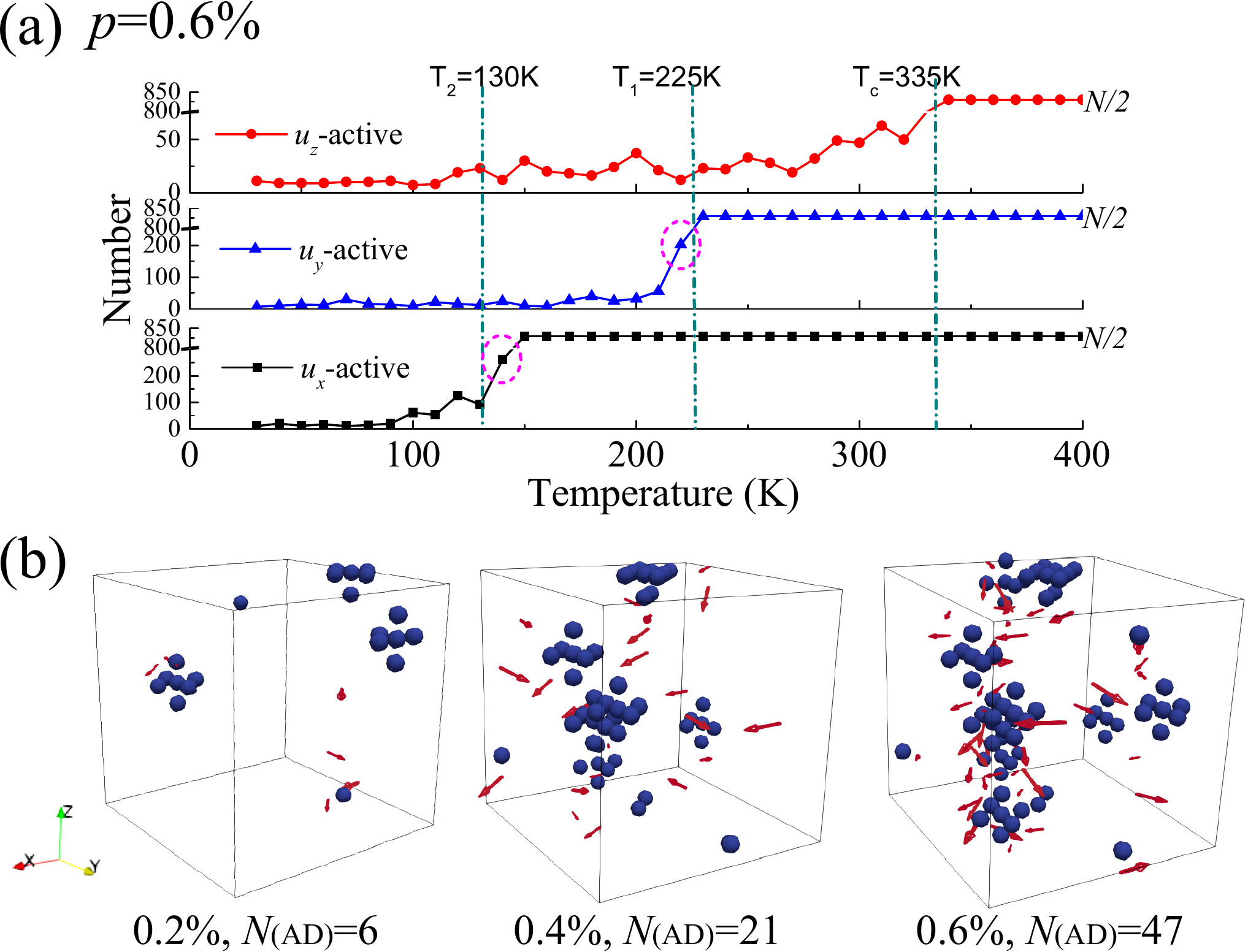} 
\par\end{centering}
\caption{(a) The evolution of active dipoles for 0.6\% BaTiO$_{3}$ and (b)
The locations and numbers of active dipoles (red arrow) and defective
dipoles (blue balls) at 300K for $p=0.2\%,\,0.4\%,\,0.6\%$. The three
curves in (a) indicate the number of active dipoles that have opposite
direction either to $\left\langle P_{x}\right\rangle $, $\left\langle P_{y}\right\rangle $,
or $\left\langle P_{z}\right\rangle $. $N_{(AD)}$ means the number
of Active dipoles. The circled points indicate notable bridging temperature
between different phases which are caused by doping. \label{fig:Active_dipoles} }
\end{figure}
Figure \ref{fig:Active_dipoles} (a) shows the number of active dipoles
as a function of temperature, which shows that even below the phase
transition temperature the number of active dipoles are not exactly
zero. For instance, at 100\,K, the BTO has the macroscopic $R3m$
phase with most of the dipoles pointing along, e.g., the $\left[111\right]$
direction. However, there are a few of the dipoles {[}see the results
labeled as $u_{x}$-active at 100\,K in Fig. \ref{fig:Active_dipoles}
(a){]} that are determined to point to the $\left[\bar{1}11\right]$
direction -- those dipoles are the active dipoles. As the temperature
reaches 200\,K, half of the dipoles become $u_{x}$-active, making
the system an $Amm2$ phase ($\left\langle P_{x}\right\rangle \simeq0$,
$\left\langle P_{y}\right\rangle =\left\langle P_{z}\right\rangle \neq0$).
In such a case, while most dipoles are along the $\left[111\right]$
and $\left[\bar{1}11\right]$ directions. However, there are dipoles
pointing along, e.g., $\left[1\bar{1}1\right]$, which are the active
dipoles {[}see the results labeled as $u_{y}$-active at 200\,K in
Fig. \ref{fig:Active_dipoles} (a){]}. At 300\,K, both the $u_{x}$-active
and $u_{y}$-active dipoles occupy half of the dipole population,
making the system an $P4mm$ phase with only $\left\langle P_{z}\right\rangle $
being nonzero. While most dipoles belong to the set of four directions
($\left[111\right]$, $\left[\bar{1}11\right]$, $\left[1\bar{1}1\right]$,
and $\left[\bar{1}\bar{1}1\right]$), the active dipoles have different
directions (e.g., $\left[11\bar{1}\right]$) {[}see the results labeled
as $u_{x}$-active at 200\,K in Fig. \ref{fig:Active_dipoles} (a){]}

When the temperature increases, the number of active dipoles initially
increases slowly at low temperatures, then at certain temperatures
it suddenly increases, finally reaching half of the whole dipole population.
Comparing to Fig. \ref{fig:Polarization_vs_Temperature}(b), it is
important to note the large jumps can be associated with the phase
change temperatures $T_{2}$, $T_{1}$ and $T_{C}$. We also find
that doped BaTiO$_{3}$ have bridging points {[}circled points in
Fig. \ref{fig:Active_dipoles}(a){]} that are hard to find in pure
BaTiO$_{3}$. The existence of such points moderate the phase transition,
causing more diffusive transition peaks observed in experiments\citep{Jin2016,Murugaraj}.

It is also critical to know how the number of active dipoles depends
on the dopant concentration. Figure \ref{fig:Active_dipoles}(b) reveals
that the number of active dipoles (red arrows) increases with dopant
concentration. More importantly, this figure shows the proximity of
active dipoles to defective dipoles, indicating a close relation between
them.

\begin{figure}[h]
\noindent \centering{}\includegraphics[width=7cm]{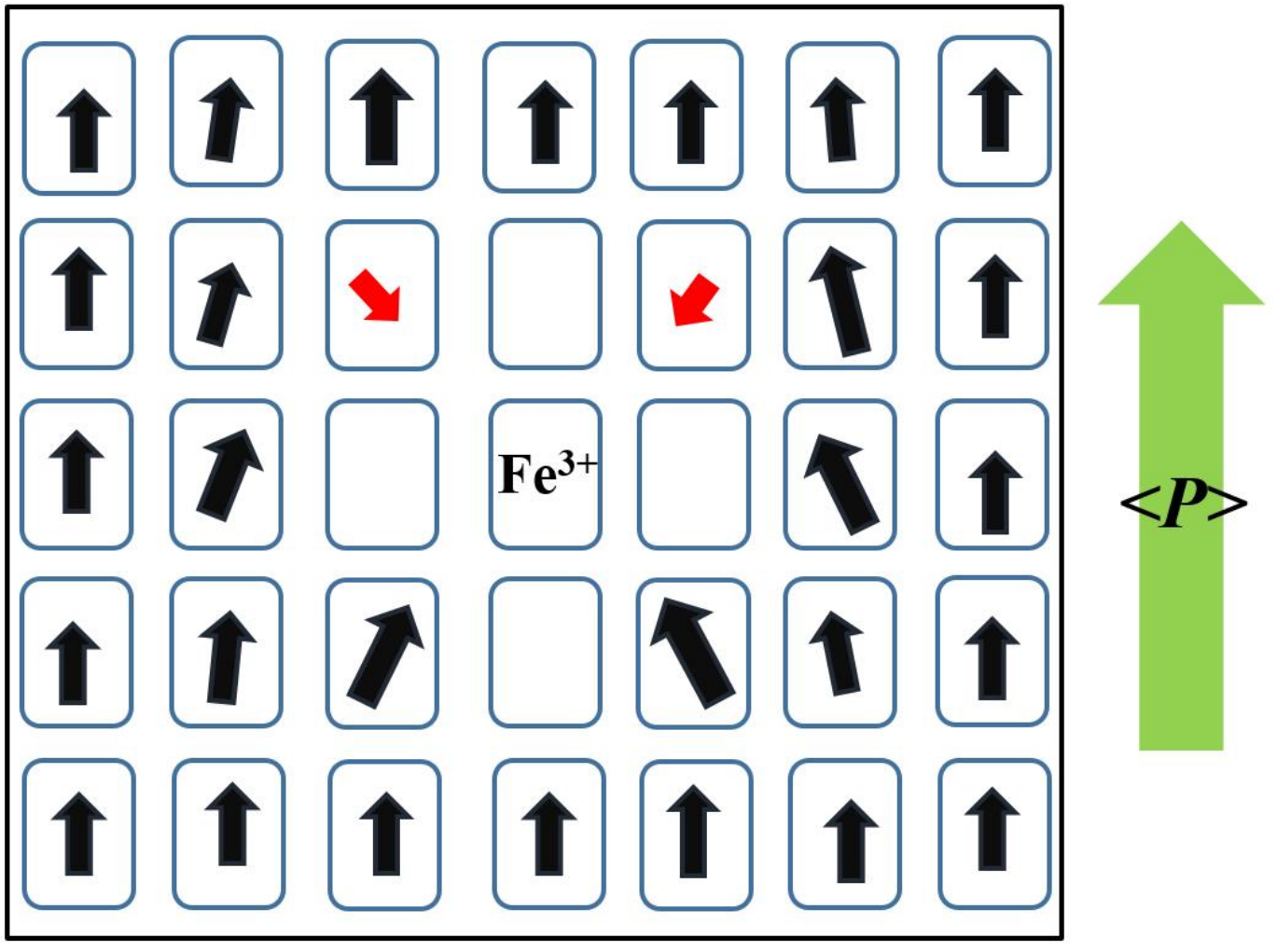}\caption{Schematic drawing for the origin of active dipoles. \label{fig:Schematic_drawing_Active_dipole}}
\end{figure}
The origin of active dipoles can be understood in terms of local chemistry
and the dipole vacuum associated with defective dipoles since (i)
Fe ions can be taken as negative when surrounded by Ti$^{4+}$ ions
(see Fig. \ref{fig:Schematic_drawing_Active_dipole}) from valence
bond theory \citep{hu_NM,defect_chemistry} and (ii) the dipole vacuum
can induce bound charges \citep{Kasap2006}. These two factors inevitably
affect the local electric field, making it different from the overall
spontaneous internal electric field, which in turn creates active
dipoles around doping sites and distorts dipole distributions.

\subsection{Hysteresis loop\label{subsec:Hysteresis-loop}}

Figure \ref{fig:Pr_Ec_value} has shown that the saturation polarization,
$P_{s}$, decreases with dopant concentration, which can be understood
with the following considerations:\textbf{ }(i) Dopant ions decrease
the total number of dipoles; (ii) Dopant ions also give rise to active
dipoles that modify the dipole distribution of pure BaTiO$_{3}$ (see
Sec. \ref{sec:Dipole-distributions} and Fig. \ref{fig:Ti_displacement});
(iii) The long-range dipole-dipole interaction is disrupted by the
defective dipoles and the induced active dipoles, so that dipoles
are not aligned as well as in pure BaTiO$_{3}$. Owing to these factors,
$P_{s}$ decreases with dopant concentration.

The remnant polarization, $P_{r}$, also shows the decreasing tendency
with doping. As a matter of fact, doping induces more active dipoles,
causing distribution variation as shown in Fig. \ref{fig:Ti_displacement}(b).
Such more symmetric distribution reduces the internal electric field
and, not surprisingly, causing $P_{r}$ to decrease. The demise of
$P_{r}$ at $p\simeq1.2\,\%$ (Fig. \ref{fig:Pr_Ec_value}) are consistent
with the distribution variation shown in Fig. \ref{fig:Ti_displacement}(b),
where doping makes the distribution peak shifts toward $u_{z}=0$,
eventually becoming a symmetric distribution at $p\simeq1.2\,\%$
where no net internal electric field is present to support $P_{r}$.

The change of the coercive field $E_{c}$ can be understood in a similar
way. As the dipole distribution becomes increasingly symmetric, the
internal electric field (associated with dipoles) becomes weaker.
Therefore a smaller external electric field can reverse the polarization,
starting from the active dipoles and eventually causing an avalanche
change of dipole direction. 

Moreover, circled points in Fig. \ref{fig:Active_dipoles}(b) indicate
that doped BaTiO$_{3}$ can have gradual changes of polarization with
temperature due to the increased number of active dipoles, which effectively
reduce the sharpness of phase transitions (see Fig. \ref{fig:Polarization_vs_Temperature}),
causing the gradual disappearance of first-order transitions (in favor
of second order phase transitions).

\subsection{Phase transition temperature\label{subsec:Phase-transition-temperature}}

In order to understand the variation of $T_{C}$, $T_{1}$, and $T_{2}$
with doping, we first note that there are two types of phase transitions
for pure and doped BaTiO$_{3}$: order-disorder and displacive. The
order-disorder phase transition happens when the correlation length
between local dipoles becomes significantly large. On the other hand,
the displacive phase transition is related to rotations of long-range
ordered dipoles. These two types of phase transitions have been discussed
in Sec. \ref{subsec:Dipole-orientation-distribution}. In addition,
it is important to note that doping in BaTiO$_{3}$ favors the $\left\langle 111\right\rangle $
and $\left\langle 110\right\rangle $ dipoles when the internal electric
field becomes weaker \citep{gamma}, which can be seen by comparing
Fig. \ref{fig:Direction_analysis}(f) to (b), which also shows that,
with 0.6\,\% doping, the number of $\left\langle 111\right\rangle $
dipoles has risen to more than $31\,\%$ of all dipoles, much higher
than the pure BaTiO$_{3}$.

Since doping introduces defective dipoles and active dipoles that
hinder the establishment of long range correlation in doped BaTiO$_{3}$,
lower $T_{C}$ is thus necessary to overcome such perturbation to
enable the order-disorder phase transition from the paraelectric phase
to the ferroelectric T phase. In addition, in the T phase, doping
makes the $u_{z}$ distribution broader and more symmetric {[}see
Fig. \ref{fig:Ti_displacement}(b){]}, which shows that the P to T
phase transition can be less dramatic, explaining the diffused phase
transition and diffusive dielectric peak seen in experiments \citep{Jin2016,Chakraborty,Hennings1982,Murugaraj}.

On the other hand, $T_{2}$ and $T_{1}$ do not decrease with doping
(in fact $T_{2}$ increases) because (i) The associated phase transitions
(e.g., the O to R phase transition) is mostly a displacive phase transition
(long range order has already been established), i.e., dipoles need
to rotate to change from the $Amm2$ phase to the $R3m$ (or $Cm$)
phase {[}see Fig. \ref{fig:Direction_analysis}(c,d) and (g,h){]}
and (ii) Doping favors the $\left\langle 111\right\rangle $ dipoles
that are building blocks of the $R3m$ or the $Cm$ phases. Therefore,
in this phase transition, the influence of doping on the formation
of long-range ordering is less relevant, while the favored $\left\langle 111\right\rangle $
dipoles help the displacive phase transition to happen, making $T_{2}$
increase with doping. The fate of $T_{1}$ can be similarly understood
by considering the relative importance of the two types of phase transitions.

From our simulations, we know that the number of defective dipoles
caused by doping plays an important role in the variation of the phase
transition temperature with dopant concentration Interestingly, different
dopants have different abilities to induce defective dipoles, largely
depending on their $3d$ electrons. For instance, Mn-doped BaTiO$_{3}$have
less defective dipoles (comparing to Fe doping) since both Mn$^{4+}$
and Mn$^{3+}$ can exist in the system and Mn$^{4+}$ is compatible
with Ti$^{4+}$ in terms of charge state \citep{Chen,Ihrig}. Therefore,
Mn-doped BaTiO$_{3}$shows less dramatic variation in the phase transition
temperature, which is consistent with experimental results \citep{Chen,Ihrig}.

\section{Conclusion \label{sec:Conclusion}}

In this work, we have developed a computationally tractable model,
which resolves around defective dipoles, to help understanding experimental
results with doped ferroelectrics. This empirical model can successfully
reproduce many important experimental results, including the ferroelectric
hysteresis loop, the phase transition temperature, and their variation
with doping. Based on the simulation results, we propose the existence
of active dipoles and show their influence on the dipole distributions,
which in turn can account for the experimentally observed phenomena.
With this approach, we are also able to correlate microscopic dipole
structural features with macroscopic phenomena. In addition, we believe
that other interpretations (e.g., defect dipole, oxygen vacancy and
local strain) also need to invoke defective dipoles before they can
explain macroscopic ferroelectric properties. Therefore, the creation
of defective dipoles (as well as active dipoles) may be seen as a
universal mechanism to account for effects associated with minuscule
doping. We thus hope that this study will help understanding and designing
novel doped perovskites to achieve desired material performance. 
\begin{acknowledgments}
We thank P. Rinke from Aalto University for fruitful discussion. This
work is financially supported by the National Natural Science Foundation
of China (NSFC), Grant No. 11574246, 51390472, U1537210, and National
Basic Research Program of China, Grant No. 2015CB654903. We also acknowledge
the ``111 Project'' of China (Grant No. B14040). L.J. acknowledges
NSFC, Grant No. 51772239 and the Fundamental Research Funds for the
Central Universities (XJTU). L.L. acknowledges NSFC, Grant No. 11564010
and the Natural Science Foundation of Guangxi, Grant No. GA139008.
D.W. also thanks the support from China Scholarship Council (201706285020). 
\end{acknowledgments}


\begin{thebibliography}{99}
\bibitem{generative} One goal of experimental work often is to find
the origin or cause, $c$, given the phenomena or evidence, $E$,
observed in an experiment. If there are several candidates as possible
causes ($c_{i}$ where $i=1,2,3\dots$), we often implicitly evaluate
the function $f\left(E\right)=\underset{c_{i}}{\textrm{arg}\textrm{max}}P\left(c_{i}|E\right),$
where $P\left(c_{i}|E\right)$ is the conditional probability that
provides the likelihood of $c_{i}$ given $E$, in order to suggest
the most likely cause of observed experimental phenomena. The calculation
of $P\left(c_{i}|E\right)$ usually is a difficult task. One way to
address this problem is to use Bayes' theorem, which states that $P\left(c_{i}|E\right)=P\left(E|c_{i}\right)P\left(c_{i}\right)/P\left(E\right)$.
Therefore the aforementioned equation can be converted to $f\left(E\right)=\underset{c_{i}}{\textrm{arg}\textrm{max}}P\left(E|c_{i}\right)P\left(c_{i}\right),$
where $P\left(E\right)$ in the denominator is removed since only
$\textrm{argmax}$ is needed. The above two expressions sometimes
are called discriminative and generative approaches in machine learning.
In this work, we show, among other things, that $P\left(E|c_{0}\right)$
is very large where $c_{0}$ represents one particular cause, i.e.,
doping can induce defective dipoles.
\bibitem{boundary} Additional tuning will be necessary to shift this
potential morphortropic boundary to room temperature for practical
applications.
\bibitem{gamma}The parameter, $\gamma$, of the effective Hamiltonian
is negative \citep{Nishimatsu}. It indicates that the energy terms
$u_{x}^{2}u_{y}^{2}$, $u_{y}^{2}u_{z}^{2}$, and $u_{x}^{2}u_{z}^{2}$
are energetically favored, which implies that $\left\langle 111\right\rangle $
and $\left\langle 110\right\rangle $ type of dipoles are more likely
than the $\left\langle 100\right\rangle $ type when only self-energy
is considered.
\end{thebibliography}
\end{document}